\documentclass[aps,pre,showpacs,twocolumn,unsortedaddress]{revtex4-1}
\usepackage{graphicx}
\usepackage{dcolumn}
\usepackage{bm}
\usepackage{amsmath}
\usepackage{mathrsfs}
\usepackage{comment}
\usepackage{color}
\begin{document}

\title{Identification of intrinsic long-range degree correlations in complex networks}

\author{Yuka Fujiki}
\email{y-fujiki@eng.hokudai.ac.jp}
\author{Kousuke Yakubo}
\email{yakubo@eng.hokudai.ac.jp} \affiliation{Department of Applied
Physics, Hokkaido University, Sapporo 060-8628, Japan}
\date{\today}

\begin{abstract}
Many real-world networks exhibit degree-degree correlations between
nodes separated by more than one step. Such long-range degree
correlations (LRDCs) can be fully described by one joint and four
conditional probability distributions with respect to degrees of two
randomly chosen nodes and shortest path distance between them. While
LRDCs are induced by nearest-neighbor degree correlations (NNDCs)
between adjacent nodes, some networks possess intrinsic LRDCs which
cannot be generated by NNDCs. Here we develop a method to extract
intrinsic LRDC in a correlated network by comparing the probability
distributions for the given network with those for nearest-neighbor
correlated random networks. We also demonstrate the utility of our
method by applying it to several real-world networks.
\end{abstract}
\pacs{89.75.Hc, 89.75.Fb, 02.70.Rr} \maketitle

\section{Introduction}
\label{sec:intro} A network representation is one of the most general
and efficient ways to describe systems consisting of elements
interacting with each other \cite{Savic19,Albert02rev,Newman03rev}. In
many networks for real-world complex systems, the number of edges from
a node, namely degree, widely fluctuates from node to node and often
obeys a power-law distribution \cite{Barabasi99}. Such broad
distributions of degrees may bring about distinctive complexity into
networks, that is, correlations between degrees of two nodes
\cite{Newman02}. Previous works on degree correlations have focused
mainly on nearest neighbor degree correlations (NNDCs) between adjacent
nodes, because NNDCs in complex networks influence their structural
features \cite{Menche10,Yook05} and dynamical properties
\cite{Schneider11,Barabasi16}. In recent years, however, it has been
elucidated that long-range degree correlations (LRDCs) beyond nearest
neighbor nodes are deeply related to various properties of networks. In
fractal complex networks like the World Wide Web, for example, highly
connected (hub) nodes repel each other over long shortest path
distances \cite{Fujiki17}. Distances between hub nodes are known to
play an important role also in the spreading of congestion
\cite{Tadic04}, the existence of an epidemic threshold \cite{Boguna13},
and other properties of functional networks \cite{Boulos13,Swanson16}.
These properties cannot be explained by NNDC. For quantifying such
long-range correlations systematically, a framework to analyze LRDCs in
complex networks has recently been formulated by introducing several
probability distributions describing LRDC \cite{Fujiki18}. By comparing
the probability distributions for a given network with those for the
corresponding random networks with the same degree distribution, one
can prove the existence of LRDC in the network and obtain detailed
information on the LRDC.

\begin{figure}[bbb]
\begin{center}
\includegraphics[width=0.45\textwidth]{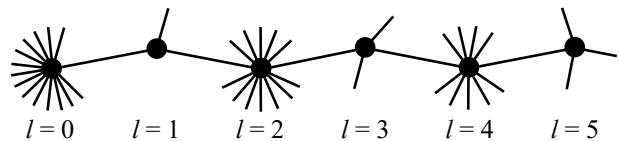}
\caption{
Schematic view of a node chain in a disassortative network. Two nodes
separated by an even (odd) distance have similar (dissimilar) degrees.
}\label{fig:1}
\end{center}
\end{figure}
It should be emphasized that NNDC generally induces LRDC. Let us
consider, for example, a network in which nodes are randomly connected
under the condition that high degree nodes are preferably connected to
low degree nodes (disassortative mixing), as shown in Fig.~\ref{fig:1}.
In this network, two nodes with similar degrees are always separated by
an even number of the shortest path distance $l$ as if
antiferromagnetic spins lying in next nearest neighbor show the
ferromagnetic order. This implies that the network shows long-range
assortative mixing at $l=2,4,\cdots$. Such an \textit{extrinsic} LRDC
induced by NNDC can also be understood from the fact that maximally
assortative or disassortative networks display community structures
\cite{Menche10}. It is impossible to judge whether observed properties
of a correlated network are caused by NNDC or \textit{intrinsic} LRDC
that cannot be produced by NNDC, if intrinsic LRDC is not segregated
from extrinsic one. In order to extract the influence of intrinsic
LRDCs on structural or dynamical properties of networks separately from
the influence of NNDCs, the identification of intrinsic LRDCs is
essential. This enables us to investigate the role of LRDCs in various
phenomena on complex networks, such as percolation, synchronization,
and epidemic spread. In the previous framework \cite{Fujiki18},
however, we cannot distinguish intrinsic LRDCs from extrinsic ones. We
provide, in this paper, a method to discriminate between extrinsic and
intrinsic LRDCs in a correlated network. To this end, we calculate the
probability distributions describing LRDCs in nearest-neighbor
correlated random networks (NNCRNs) which are maximally randomized
networks under the constraint that NNDC is specified. One can identify
intrinsic LRDC in a network by comparing the probability distributions
with those for its corresponding NNCRNs having the same NNDC as the
original network. We also demonstrate the utility of our method by
applying it to several real-world networks.

The rest of this paper is organized as follows. In
Sec.~\ref{sec:framework}, we briefly summarize the previous work
\cite{Fujiki18} regarding the general description of LRDCs. Examples of
extrinsic LRDCs induced by NNDCs are clearly shown by means of the
probability distributions. In Sec.~\ref{sec:nncrn}, we present the
mean-field calculation of the probability distributions for NNCRNs in
order to extract intrinsic LRDCs. We demonstrate, in
Sec.~\ref{sec:demonstration}, how to interpret the observed LRDCs by
applying our method to several real-world networks. Finally, we
conclude in Sec.~\ref{sec:conclusion}.

\section{Identification of long-range degree correlations}
\label{sec:framework} In this section, we briefly explain the way to
describe LRDCs and to judge the existence of them in complex networks,
according to the previous work \cite{Fujiki18}. We also demonstrate
that NNDC inevitably induces LRDC even in a network with no intrinsic
LRDCs.

\subsection{Description of LRDC}
\label{subsec:description}

A general pairwise LRDC can be described by the joint probability
distribution $P(k,k',l)$ that two randomly chosen nodes have degrees
$k$ and $k'$ and the shortest path distance between them is $l$. Let us
assume that $P(k,k',l)$ is defined also for pairs of identical nodes
($l=0$) and for disconnected node pairs. Thus, $P(k,k',l)$ is
normalized as $\sum_{k,k',l}P(k,k',l)=1$, where the summation over $l$
includes $l=0$ and the virtual distance (denoted by $l_{\infty}$)
between disconnected pairs. If the distribution $\tilde{P}(k,k',l)$
defined only for connected pairs is required, one can construct it from
$P(k,k',l)$ as
$\tilde{P}(k,k',l)=P(k,k',l)/\sum_{k,k'}\sum_{l=1}'P(k,k',l)$, where
$\sum_{l}'$ does not include $l=l_{\infty}$. In addition to the joint
probability, it is convenient to introduce four conditional probability
distributions $P(l|k,k')$, $P(k'|k,l)$, $P(k,k'|l)$, and $P(k',l|k)$
which are defined by
\begin{subequations}
\begin{align}
P(l|k,k') &= \frac{P(k,k',l)}{\sum_{l}P(k,k',l)}, \label{def_pl_kk}\\
P(k'|k,l) &= \frac{P(k,k',l)}{\sum_{k'}P(k,k',l)}, \label{def_pk_kl}\\
P(k,k'|l) &= \frac{P(k,k',l)}{\sum_{k,k'}P(k,k',l)}, \label{def_pkk_l}\\
P(k',l|k) &= \frac{P(k,k',l)}{\sum_{k',l}P(k,k',l)}. \label{def_pkl_k}
\end{align}
\label{def_cond_p}
\end{subequations}

The meanings of these probability distributions are obvious. The
probability $P(l|k,k')$, for example, is the shortest path distance
distribution between nodes of degrees $k$ and $k'$, $P(k'|k,l)$ is the
degree distribution of a node separated by $l$ from a node of degree
$k$, and so on. LRDC in a network can be characterized by one of these
five probability distributions. We should note that $P(k,k',l)$,
$P(l|k,k')$, and $P(k',l|k)$ are meaningless for networks with
infinitely large components, because these distributions are always
zero for finite $l$.

The above probability distributions also describe degree correlation
between nearest-neighbor nodes. It is well understood that NNDC is
captured by the joint probability $P_{\text{nn}}(k,k')$ that two end
nodes of a randomly chosen edge have the degrees $k$ and $k'$ or the
conditional probability $P_{\text{nn}}(k'|k)$ that a node adjacent to a
randomly chosen node of degree $k$ has the degree $k'$
\cite{Maslov02,Pastor-Satorras01}. These probability distributions
$P_{\text{nn}}(k,k')$ and $P_{\text{nn}}(k'|k)$ are equivalent to
$P(k,k'|l=1)$ and $P(k'|k,l=1)$, respectively. In this context, the
probability distributions describing LRDC are a natural extension of
$P_{\text{nn}}(k,k')$ and $P_{\text{nn}}(k'|k)$. In fact, the
well-known relation $P_{\text{nn}}(k'|k)=
P_{\text{nn}}(k,k')/\sum_{k'}P_{\text{nn}}(k,k')$ is a special case of
$P(k'|k,l)=P(k,k'|l)/\sum_{k'}P(k,k'|l)$.

Using the joint and conditional probability distributions, we can
define various useful indices reflecting specific aspects of LRDC. For
example, the average degree of the $l$th neighbor nodes of a node of
degree $k$ is defined by
\begin{align}\label{klk}
k_l (k) =\sum_{k'}k'P(k'|k,l),
\end{align}
which is an extension of the average degree of neighbors of a node of
degree $k$, $k_{\text{nn}}(k)=\sum_{k'}k'P_{\text{nn}}(k'|k)$
\cite{Echenique05}. Another example is the average shortest path
distance between two nodes of degrees $k$ and $k'$, which is defined by
\begin{equation}
\langle l (k,k') \rangle = \sum_{l}l\tilde{P}(l|k,k'),
\label{lavkk}
\end{equation}
where $\tilde{P}(l|k,k')=P(l|k,k')/\sum_{l=1}'P(l|k,k')$ is the
distribution defined only for connected pairs. By means of this index,
we can discuss long-range repulsive or attractive correlation between
hub nodes. Besides the above, many indices characterizing NNDCs, such
as assortativity \cite{Newman02}, can be extended for LRDCs
\cite{Mayo15,Arcagni17}.

\subsection{Existence of LRDC}
\label{subsec:existence}

A network has LRDC if the probability distributions for the network are
different from those for a \textit{long-range uncorrelated network}
with the same degree distribution. In the case of nearest-neighbor
uncorrelated networks, the joint probability  satisfies
$P_{\text{nn}}(k,k')=Q_{\text{nn}}(k)Q_{\text{nn}}(k')$, where
$Q_{\text{nn}}(k)=kP(k)/\langle k\rangle$ is the probability that one
end node of a randomly chosen edge has the degree $k$, $P(k)$ is the
degree distribution, and $\langle k\rangle$ is the average degree.
Analogously, a long-range uncorrelated network is considered to be
defined as a network for which the relation
\begin{equation}
P(k,k'|l) =Q(k|l)Q(k'|l) ,
\label{def_un}
\end{equation}
is satisfied for any $l$, where $Q(k|l)=\sum_{k'}{P(k,k'|l)}$ is the
probability that one node of a randomly chosen node pair separated by
$l$ from each other has the degree $k$. The condition of
Eq.~(\ref{def_un}) can be established if the network size is infinite,
while $P(k,k',l)$, $P(l|k,k')$, and $P(k',l|k)$ are meaningless in such
networks. In finite networks, in contrast, the finite-size effect
prevents to hold Eq.~(\ref{def_un}) even in random networks. It is,
however, rather practically important to compare LRDC in a given
network with that in a random network with the same size and the same
degree distribution. Hereafter, a network $G$ is regarded to have LRDC
if the probability distributions for $G$ differ from those for the
corresponding random network. We denote the probability distribution
functions for random networks by adding the subscript $0$.

One can analytically calculate the five probability distributions for a
random network with a given size $N$ and a given degree distribution
$P(k)$ within the mean-field approximation \cite{Fujiki18,Melnik16}. In
this theory, the shortest path distance distribution $P_{0}(l|k,k')$
between nodes of degrees $k$ and $k'$ is first computed, then the joint
probability distribution $P_{0}(k,k',l)$ is calculated from the general
relation $P_{0}(k,k',l)=P(k)P(k')P_{0}(l|k,k')$. Furthermore,
Eq.~(\ref{def_cond_p}) presents the other three conditional probability
distributions. Details of the calculations will be explained in
Sec.~\ref{sec:nncrn}. The accuracy of the above theoretical calculation
is basically quite high, while it becomes relatively poor if the random
network is very close to the percolation transition because the
mean-field theory assumes a narrow distribution of component sizes. The
existence of LRDC in a network $G$ can be proven by comparing
$P(k,k',l)$ (or other distributions) for $G$ with $P_{0}(k,k',l)$
calculated theoretically for the corresponding random network with the
same $N$ and the same $P(k)$ as $G$.

\subsection{Extrinsic LRDC}
\label{sec:extrinsic}

Within the above framework, it has been revealed that many real-world
networks exhibit LRDCs \cite{Fujiki18}. As mentioned in
Sec.~\ref{sec:intro}, observed degree correlations, however, include
extrinsic LRDCs which are purely generated by NNDCs. Here we
demonstrate numerically how NNDCs induce LRDCs. For this purpose, we
prepare nearest-neighbor correlated random networks (NNCRNs) in which
nodes are randomly connected while a specific NNDC is preserved. More
specifically, we first rewire, referring to the procedure adopted in
\cite{Menche10}, edges in an Erd\H{o}s-R\'{e}nyi random graph to
increase or decrease the Spearman's rank correlation coefficient
$\varrho$ defined by \cite{Litvak13}
\begin{equation}
\varrho=\frac{\sum_{k,k'}R_{k}R_{k'}P_{\text{nn}}(k,k')-\left[\sum_{k}R_{k}Q_{\text{nn}}(k) \right]^{2}}
{\sum_{k}R_{k}^{2}Q_{\text{nn}}(k)-\left[\sum_{k}R_{k}Q_{\text{nn}}(k) \right]^{2}},
\label{eq:spearman}
\end{equation}
where $R_{k}$ given by \cite{Zhang16}
\begin{equation}
R_{k}=\frac{NkP(k) +1}{2}+N\sum_{k'=0}^{k-1}k'P(k'),
\label{eq:rank}
\end{equation}
is the rank of the degree $k$. The correlation coefficient $\varrho$
has a similar meaning to assortativity $r$ \cite{Newman02}, but does
work well even for heterogeneous networks with the diverging
third-order moment of the degree distribution, unlike the case of $r$.
When $\varrho$ becomes a desired value, the assortative/disassortative
rewiring operation is switched to a random rewiring procedure while
keeping $P_{\text{nn}}(k,k')$. After a sufficient number of random
rewiring operations, we have an assortative/disassortative NNCRN with
the same degree distribution as the initial Erd\H{o}s-R\'{e}nyi random
graph.

\begin{figure*}[ttt!]
\begin{center}
\includegraphics[width=0.9\textwidth]{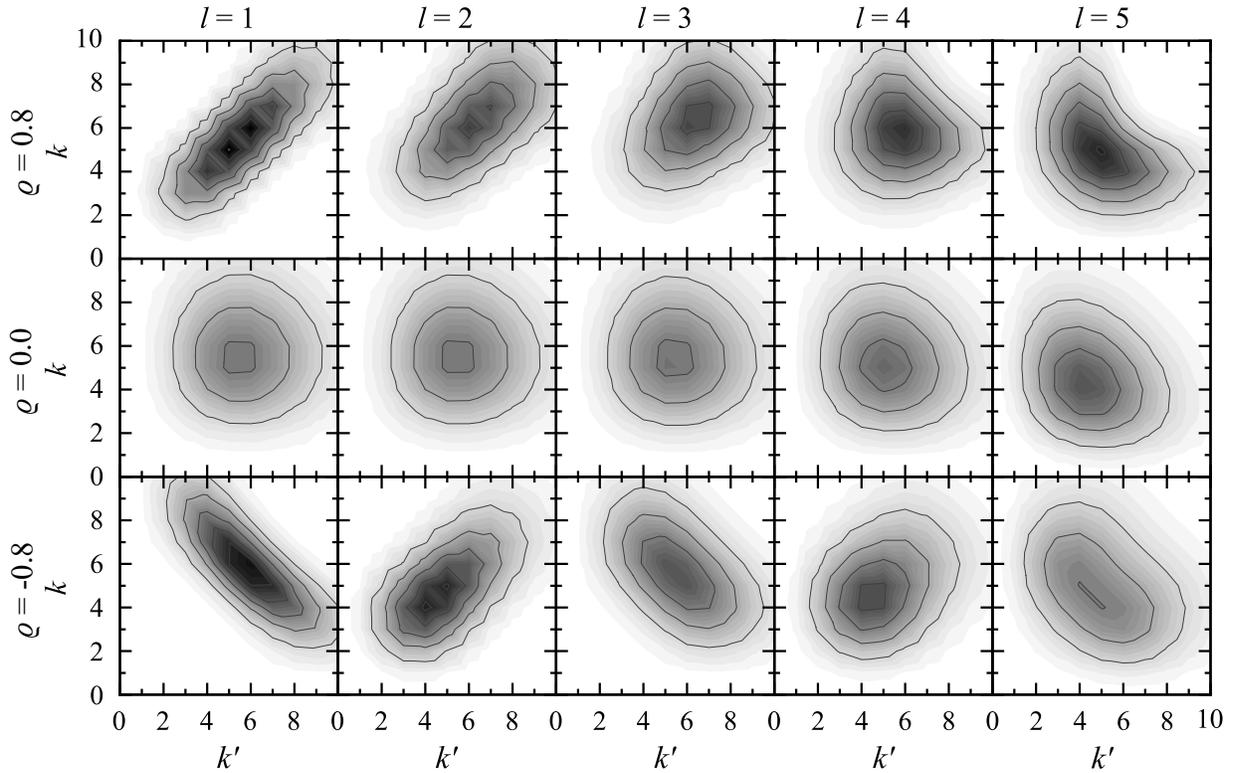}
\caption{
Contour plots of the probability distribution $P(k,k'|l)$ for
NNCRNs with $\varrho=0.8$ (upper panels, $r=0.76$),
$0.0$ (middle panels, $r=0.0$),
and $-0.8$ (lower panels, $r=-0.75$)
as functions of $k$ and $k'$ for
$l=1$ to $5$,
where $r$ is the assortativity \cite{Newman02}.
The degree sequence is the same as that of
Erd\H{o}s-R\'{e}nyi random graphs with $N=1,000$ and
$\langle k\rangle=5.0$.
}\label{fig:2}
\end{center}
\end{figure*}
Figure \ref{fig:2} represents the probability distribution $P(k,k'|l)$
measured for NNCRNs which are numerically prepared by the above method.
The network size is $N=1,000$ and the average degree is set as
$\langle k\rangle=5.0$. The distribution $P(k,k'|l)$ is averaged over
$100$ samples. The upper five panels show the results for the assortative
NNCRN with $\varrho=0.8$ as functions of $k$ and $k'$ for $l=1$ to $5$
(from left to right). The distribution $P(k,k'|l=1)$ clearly indicates
a strong positive correlation between $k$ and $k'$. This tendency is
found up to about $l=3$ but becomes weaker as $l$ increases. For $l=5$,
in contrast, $P(k,k'|l)$ displays weak negative degree correlation
caused by the reaction of assortative mixing for $l\le 3$. The results
for the disassortative NNCRN with $\varrho=-0.8$ are shown in the lower
panels. These results demonstrate that the sign of the correlation
between $k$ and $k'$ alternately changes as $l$ is incremented. This
behavior of $P(k,k'|l)$ can be easily understood from Fig.~\ref{fig:1}.
Such an alternative correlation-sign change in nearest-neighbor
correlated orders can also be seen in an antiferromagnetic spin
ordering and a color ordering of graph coloring for tree-like graphs.
Assortative and disassortative degree correlations shown in
Fig.~\ref{fig:2} for $l\ge 2$ are purely induced by NNDCs, namely,
extrinsic LRDCs. The purpose of this work is to distinguish between
extrinsic and intrinsic LRDCs in a network with both types of
correlations.

\section{Mean-field analysis}
\label{sec:nncrn}

As LRDC in a given network is detected by comparing its probability
distributions with those for the corresponding random network, we can
identify intrinsic LRDC by comparing with those for the corresponding
NNCRN with the same size and the same NNDC. In this section, we
analytically calculate the probability distributions for an NNCRN with
a given size $N$ and a given $P_{\text{nn}}(k,k')$ within the
mean-field approximation and the local-tree approximation. Hereafter,
we denote the probability distribution functions for NNCRNs by adding
the subscript $1$.

We first focus on the shortest path distance distribution
$P_{1}(l|k,k')$ between nodes of degrees $k$ and $k'$. Such a
distribution for a connected NNCRN has been calculated by Melnik and
Gleeson \cite{Melnik16}. The distribution $P_{1}(l|k,k')$ for a network
with multiple components can be calculated by extending their argument.
As mentioned in Sec.~\ref{subsec:description}, $P_{1}(l|k,k')$ is
normalized with respect to $l$ including $l=0$ and the virtual distance
($l_{\infty}$) between disconnected pairs. In order to calculate
$P_{1}(l|k,k')$ both for connected and disconnected node pairs, let us
introduce the cumulative probability
$\rho(l|k,k')=\sum^{l}_{l'=0}P_{1}(l'|k,k')$ only for
$l\ne l_{\infty}$. This probability is expressed by
\begin{equation}
\rho(l|k,k')=1-\left[1-\frac{\delta_{kk'}}{NP(k)}\right]\left[1-\bar{q}(l-1|k,k')\right]^{k} ,
\label{cal_rho}
\end{equation}
where $\bar{q}(l|k,k')$ is the probability that an adjacent node of a
randomly chosen node $i_{k}$ of degree $k$ lies within the distance $l$
from a specific node $j_{k'}$ of degree $k'$, and $\bar{q}(-1|k,k')=0$.
The prefactor $[1-\delta_{kk'}/NP(k)]$ represents the probability that
the node $i_{k}$ does not coincide with the node $j_{k'}$. Thus, the
right-hand side of Eq.~(\ref{cal_rho}) is the probability that at least
one of $k$ adjacent nodes of $i_{k}(\ne j_{k'})$ lies within the
distance $l-1$ from $j_{k'}$. From the definition, the probability
$\bar{q}(l|k,k')$ is given by \cite{Melnik16}
\begin{equation}
\bar{q}(l|k,k')=\sum_{k''}P_{\text{nn}}(k''|k)q(l|k'',k') ,
\label{rel_qbar_q_1}
\end{equation}
where $q(l|k,k')$ is the probability that a terminal node $i_{k}$ of a
randomly chosen edge lies within the distance $l$ from a node $j_{k'}$
of degree $k'$ given that $i_{k}$ has the degree $k$ and is closer to
$j_{k'}$ than the other terminal node of the edge. We notice here that
$P(k)$ in Eq.~(\ref{cal_rho}) and $P_{\text{nn}}(k'|k)$ in
Eq.~(\ref{rel_qbar_q_1}) are presented by the specified
$P_{\text{nn}}(k,k')$ as $P(k)=\langle
k\rangle\sum_{k'}P_{\text{nn}}(k,k')/k$ and
$P_{\text{nn}}(k'|k)=\langle k\rangle P_{\text{nn}}(k,k')/kP(k)$,
respectively. A similar idea to Eq.~(\ref{cal_rho}) leads to another
relation between $\bar{q}(l|k,k')$ and $q(l|k,k')$, which is given by
\begin{equation}
q(l|k,k')=1-\left[1-\frac{\delta_{kk'}}{NP(k)}\right]\left[1-\bar{q}(l-1|k,k')\right]^{k-1} .
\label{rel_qbar_q_2}
\end{equation}
We can iteratively solve the coupled recurrence equations
(\ref{rel_qbar_q_1}) and (\ref{rel_qbar_q_2}) with the initial
condition,
\begin{equation}
\bar{q} (0|k,k') =\frac{P_{\mathrm{nn}} (k'|k)}{NP(k')},
\label{ini_cond2}
\end{equation}
which is the probability that an adjacent node of $i_{k}$ is the
specific node $j_{k'}$ itself. The cumulative probability
$\rho(l|k,k')$ is obtained by substituting the solution to
Eq.~(\ref{cal_rho}).

\begin{figure*}[t!]
\begin{center}
\includegraphics[width=0.95\textwidth]{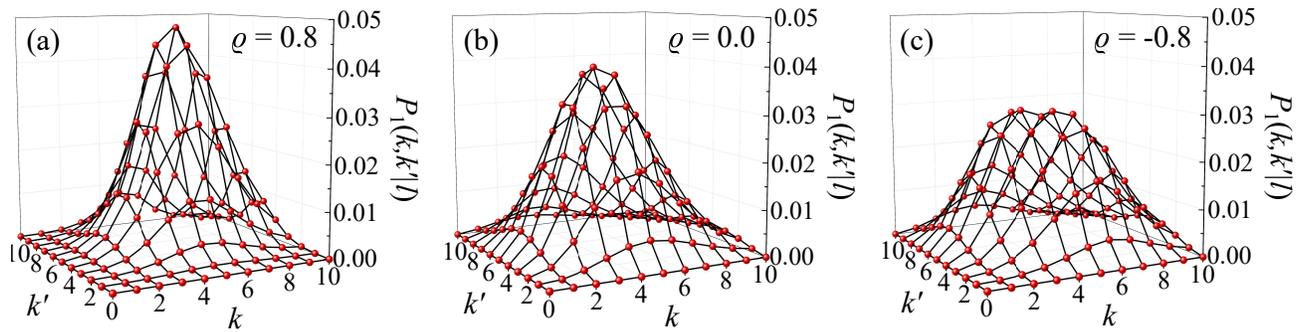}
\caption{(Color online)
Conditional probability distributions $P_{1}(k,k'|l=5)$ as functions of
$k$ and $k'$ for assortative/disassortative NNCRNs treated in
Fig.~\ref{fig:2}. Red symbols represent the numerical results which are
the same as the results for $l=5$ shown in Fig.~\ref{fig:2}, and the
wireframes indicate $P_{1}(k,k'|l=5)$ calculated in the mean-field
approximation. The Spearman's rank correlation coefficients are (a)
$\varrho=0.8$, (b) $\varrho=0.0$, and (c) $\varrho=-0.8$.
}\label{fig:3}
\end{center}
\end{figure*}
Although the cumulative probability $\rho(l|k,k')$ must be symmetric
with respect to $k$ and $k'$ from the definition, $\rho(l|k,k')$ given
by Eq.~(\ref{cal_rho}) is obviously asymmetric. This is due to the
asymmetric treatment of the source node $j_{k'}$ and sink node $i_{k}$,
and influences the accuracy of the mean-field approximation. For
$k'>k$, $\rho(l|k,k')$ is generally more accurate than $\rho(l|k',k)$
\cite{Fujiki18}. To improve the accuracy, we first calculate
$\rho(l|k,k')$ for $k<k'$ by Eq.~(\ref{cal_rho}), then transfer it to
$\rho(l|k',k)$. Using the symmetrized $\rho(l|k,k')$, the probability
distribution $P_{1}(l|k,k')$ is then calculated by
\begin{equation}
P_{1}(l|k,k')=\rho(l|k,k')-\rho(l-1|k,k'),
\label{rho_to_p1}
\end{equation}
where $P_{1}(0|k,k')=\rho(0|k,k')=\delta_{kk'}/NP(k)$, and for
disconnected node pairs $P_{1}(l_{\infty}|k,k')=1- \lim_{l\to
\infty}\rho(l|k,k')$. The joint probability distribution
$P_{1}(k,k',l)$ is calculated from the general relation
(\ref{def_pl_kk}) and $\sum_lP(k,k',l)=P(k)P(k')$, which is given by
\begin{equation}
P_{1}(k,k',l)=P(k)P(k')P_{1}(l|k,k') .
\label{rel_qbar_joint}
\end{equation}
Other three conditional probability distributions $P_{1}(k'|k,l)$,
$P_{1}(k,k'|l)$, and $P_{1}(k',l|k)$ are computed by
Eq.~(\ref{def_cond_p}).

We have checked the accuracy of the above mean-field approximation. To
this end, the same networks as in Fig.~\ref{fig:2} have been treated.
Figure \ref{fig:3} shows $P(k,k'|l=5)$ for (a) $\varrho=0.8$, (b)
$\varrho=0.0$, and (c) $\varrho=-0.8$. Black wireframes obtained by the
mean-field calculation agree quite well with red symbols indicating the
corresponding numerical results. Since the probability distributions
other than $P(k,k'|l)$ are rigorously related to $P(k,k'|l)$ through
Eq.~(\ref{def_cond_p}), the accuracy of the mean-field approximation
for these distributions including the joint probability distribution
are the same with $P(k,k'|l)$.

\begin{table}[bbb]
\caption{\label{tab:table1}
Maximum values $d_{\text{max}}$ of
$d(P_{1}^{\text{ana}},P_{1}^{\text{num}})$ as a function of the average
degree for NNCRNs with specific values of $\varrho$ and the same degree
distribution as Erd\H{o}s-R\'{e}nyi random graphs, the average degree
$\langle k\rangle_{\text{max}}$ giving $d_{\text{max}}$, and the
critical average degree $\langle k\rangle_{\text{c}}$ for the
percolation transition. $\langle k\rangle_{\text{c}}$'s for
$\varrho=0.8$ and $-0,8$ have been estimated numerically by evaluating
the largest component size. For $\varrho=0.0$, $\langle
k\rangle_{\text{c}}$ is known to be unity \cite{Albert02rev}.}
 \begin{ruledtabular}
  \begin{tabular}{lccc}
                                      & $\varrho=0.8$ & $\varrho=0.0$ & $\varrho=-0.8$ \\ \hline
    $d_{\text{max}}$                  & $0.66$        & $0.66$        & $0.57$         \\
    $\langle k\rangle_{\text{max}}$   & $1.05$        & $1.10$        & $1.25$         \\
    $\langle k\rangle_{\text{c}}$     & $0.8$         & $1.0$         & $1.1$          \\
  \end{tabular}
 \end{ruledtabular}
\label{table:1}
\end{table}
\begin{figure*}[ttt!]
\begin{center}
\includegraphics[width=1.0\textwidth]{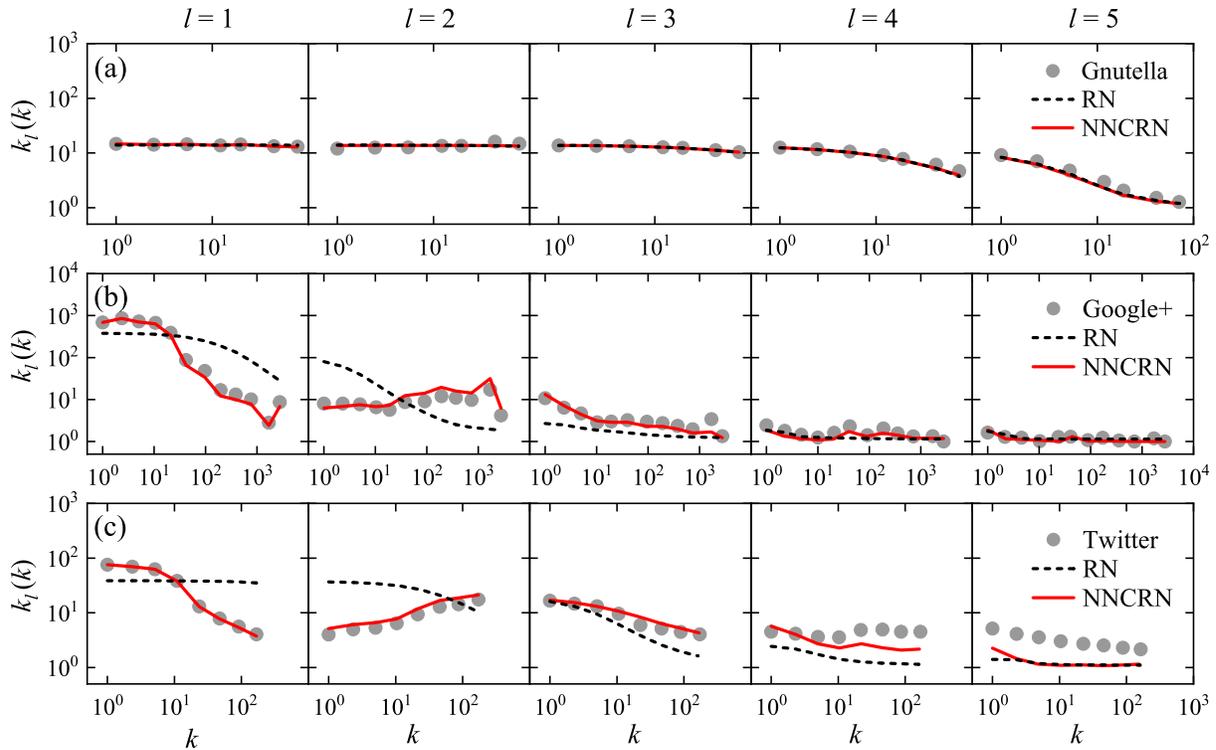}
\caption{(Color online)
Average degree $k_{l}(k)$ of the $l$th neighbor nodes of a node of
degree $k$ for (a) the Gnutella network, (b) Google+ network, and (c)
Twitter network. Gray symbols represent $k_{l}(k)$ for the real-world
networks at fixed values of $l$. Dashed black lines indicate $k_{l}(k)$
for the corresponding random networks with the same degree
distributions as the real-world networks. Solid red
lines are for the NNCRNs with the same NNDCs.}
\label{fig:4}
\end{center}
\end{figure*}
In order to quantify the accuracy of the mean-field approximation, we
introduce the distribution distance defined by
\begin{equation}
d(P,P')=\sqrt{\frac{\sum_{k,k'}\sum_{l=1}'\left[P(k,k',l)-P'(k,k',l)\right]^2}
{\sum_{k,k'}\sum_{l=1}'\left[P(k,k',l)^{2}+P'(k,k',l)^{2}\right]}},
\label{distance}
\end{equation}
where $P(k,k',l)$ and $P'(k,k',l)$ are two joint probability
distributions to be compared and the summation $\sum_{l}'$ does not
include $l=l_{\infty}$. This quantity is bounded as
$0\le d(P,P')\le 1$, where $d(P,P')=0$ for $P(k,k',l)=P'(k,k',l)$
and $d(P,P')=1$ when these two distributions do not overlap each
other at all. We have computed the distribution distance
$d(P_{1}^{\text{ana}},P_{1}^{\text{num}})$ between two distributions
$P_{1}^{\text{ana}}(k,k',l)$ and $P_{1}^{\text{num}}(k,k',l)$
calculated analytically and numerically, respectively, for NNCRNs used
in Fig.~\ref{fig:3}. The results are
$d(P_{1}^{\text{ana}},P_{1}^{\text{num}})=0.027$, $0.016$, and $0.013$
for $\varrho=0.8$, $0.0$, and $-0.8$, respectively. These distribution
distances very closed to zero quantitatively indicate the high accuracy
of the analytical calculation. It should be noted, however, that the
mean-field approximation has poor accuracy if an NNCRN is very close to
its percolation transition point. This can be confirmed by the fact
that the distribution distance
$d(P_{1}^{\text{ana}},P_{1}^{\text{num}})$ as a function of the average
degree $\langle k\rangle$ displays a sharp peak and takes the maximum
value  near (or slightly larger than) the critical point
$\langle k\rangle_{\text{c}}$ as shown by Table \ref{table:1}. This
is because the mean-field treatment ignores the fluctuation of the size
of the component containing the source node $j_{k'}$, while the
component-size distribution is very wide near the percolation threshold.

\section{Real-world networks}
\label{sec:demonstration} In this section, we demonstrate how detailed
information on LRDC in a given network can be obtained from the
probability distributions. For this purpose, we first analyze
contrasting three networks; the Gnutella network \cite{url2}, Google+
network \cite{url1}, and Twitter network \cite{url1}. The Gnutella
network ($\langle{k}\rangle=7.4$) represents the connection between
$10,876$ hosts sharing files by peer-to-peer. The Google+ network
($\langle{k}\rangle=3.3$) represents the friendship relation between
$23,628$ users of Google+. The Twitter network
($\langle{k}\rangle=2.8$) is comprised of the follower/following
relationship between $23,370$ accounts. The average shortest path
distances are $\langle l\rangle=4.64$, $4.03$, and $6.30$ for the
Gnutella, Google+, and Twitter networks, respectively. In analyses of
these networks, weights and directions of edges are ignored.

Symbols in Fig.~\ref{fig:4} shows the $k$ dependence of the average
degree of the $l$th neighbor nodes $k_{l}(k)$ defined by
Eq.~(\ref{klk}) for these networks. The probability distribution
$P(k'|k,l)$ is measured directly from the network data. Dashed black
lines and red solid lines indicate $k_{l}(k)$ for corresponding random
networks with the same degree distributions and NNCRNs with the same
NNDCs obtained from $P_{0}(k'|k,l)$ and $P_{1}(k'|k,l)$, respectively.
The probability distributions $P_{0}(k'|k,l)$ and $P_{1}(k'|k,l)$ are
calculated analytically by using $P(k)$ and $P_{\text{nn}}(k'|k)$ of
the original networks. Gray symbols for the Gnutella network coincide
with the dashed black and solid red lines for any $l$ from $l=1$ to
$5$, which means that the Gnutella network is totally uncorrelated. In
contrast, the results for the Google+ and the Twitter networks deviate
from $k_{l}(k)$ for their corresponding random networks (dashed black
line) at least for $l\le 3$. This implies that these networks exhibit
LRDCs. The result for the Google+ network agrees well with that for the
corresponding NNCRN (solid red line) for any $l$. Therefore, we can
conclude that the LRDC in the Google+ network is induced by an NNDC,
i.e., the LRDC is extrinsic. On the other hand, $k_{l}(k)$ for the
Twitter network differs from both the dashed black and solid red lines
for $l\ge 4$, which indicates that the LRDC in the Twitter network is
intrinsic. As illustrated by these examples, real-world networks are
classified into three types: uncorrelated networks, only
nearest-neighbor degree correlated networks (extrinsically long-range
correlated networks), and intrinsically long-range correlated networks.

\begin{table}[bbb!]
\caption{\label{table:2}
Characteristics of LRDCs in real-world complex networks. $N$,
$\langle k\rangle$, and  $\varrho$ are the network size, average degree,
and Spearman's rank correlation coefficients, respectively. The quantities
$d_{0}$ and $d_{1}$ represent the values of $d(P,P_{0})$ and $d(P,P_{1})$,
respectively.}
\begin{tabular}{lcccccc}
\hline\hline
Network                &$N$       &$\langle{k}\rangle$  &$\varrho$  &$d_0$  &$d_1$  &Ref.\\
\hline
Gnutella               &10,876    &7.4   &0.00  &0.11 &0.13 &\cite{url2}\\
Internet (AS level) (1)&10,515    &4.1   &-0.55 &0.17 &0.25 &\cite{url4}\\
Internet (AS level) (2)&26,475    &4.0   &-0.53 &0.21 &0.25 &\cite{url1}\\
Internet (AS level) (3)&22,963    &4.2   &-0.51 &0.19 &0.26 &\cite{url3}\\
Google+                &23,628    &3.3   &-0.74 &0.63 &0.31 &\cite{url1}\\
Email                  &36,692    &10.0  &-0.01 &0.37 &0.40 &\cite{url2}\\
WWW (1)                &415,624   &11.4  &-0.32 &0.87 &0.41 &\cite{url1}\\
Brightkite             &58,228    &7.4   &0.22  &0.54 &0.41 &\cite{url1}\\
Facebook               &63,731    &25.6  &0.38  &0.73 &0.43 &\cite{url1}\\
Coauthor (1)           &18,771    &21.1  &0.35  &0.74 &0.48 &\cite{url1}\\
YouTube                &1,134,890 &5.3   &-0.08 &0.63 &0.50 &\cite{url1}\\
Coauthor (2)           &12,006    &19.7  &0.72  &0.89 &0.52 &\cite{url2}\\
Actor                  &82,583    &88.8  &0.34  &0.73 &0.57 &\cite{url4}\\
Coauthor (3)           &23,133    &8.1   &0.26  &0.71 &0.61 &\cite{url2}\\
Twitter                &23,370    &2.8   &-0.74 &0.82 &0.67 &\cite{url1}\\
WWW (2)                &325,729   &6.7   &-0.11 &0.90 &0.81 &\cite{url1}\\
Internet (router level)&192244    &6.3   &0.27  &0.89 &0.82 &\cite{url5}\\
WWW (3)                &255,265   &15.2  &-0.20 &0.95 &0.88 &\cite{url1}\\
WWW (4)                &685,230   &19.4  &-0.17 &0.96 &0.92 &\cite{url1}\\
Protein folding        &132,167   &3.5   &0.51  &0.97 &0.95 &\cite{url4}\\
Amazon                 &334,863   &5.5   &-0.19 &0.96 &0.96 &\cite{url1}\\
\hline\hline
\end{tabular}
\end{table}
We have examined quantitatively LRDCs in many real-world networks by
utilizing the distribution distance $d(P,P')$ defined by
Eq.~(\ref{distance}). The strength of LRDC in a given network $G$ is
quantified by the distance $d(P,P_{0})$ between the joint probability
distributions $P(k,k',l)$ for $G$ and $P_{0}(k,k',l)$ for the
corresponding random network with the same degree distribution as $G$.
In addition, the strength of the intrinsic character of LRDC in the
network $G$ is measured by the distance $d(P,P_{1})$ between
$P(k,k',l)$ and $P_{1}(k,k',l)$ for the NNCRN corresponding to $G$. The
quantity $d(P,P_{1})$ becomes zero or unity if the LRDC in $G$ is
completely extrinsic or intrinsic, respectively. The results for $21$
real-world networks are summarized in Table \ref{table:2}. Except for
the first four networks with low values of $d(P,P_{0})$, LRDCs in many
real-world networks present a strong intrinsic character with large
values of $d(P,P_{1})$. It is interesting to notice that networks
showing large values of $d(P,P_{0})$ give large $d(P,P_{1})$'s. This
means that LRDCs in strongly correlated real networks are dominated by
intrinsic ones.

It should be noted that three AS-level Internet graphs show strongly
disassortative mixing with large negative $\varrho$ while $d(P,P_{0})$
is close to zero. An uncorrelated network with $d(P,P_{0})=0$ can show
disassortative mixing if the degree distribution is highly
inhomogeneous, which is known as structural disassortativity
\cite{Maslov04}. For example, a randomly connected network with the
degree distribution $P(k)=[\delta_{k,N-1}+(N-1)\delta_{k,1}]/N$
(star-graph degree distribution) has $\varrho=-1$ independently of the
network size $N$. In fact, the largest degree node in the network
``Internet (AS level) (1)" is connected to more than $20$\% of all
nodes ($N=10,515$ and $k_{\text{max}}=2,277$). For identifying
essential degree correlation caused by biased interactions in
distinction from apparent structural disassortativity due to
inhomogeneous degree distribution, we need to investigate the
difference between $P(k,k',l)$ and $P_{0}(k,k',l)$ in addition to an
assortativity measure like $\varrho$.

\begin{figure*}[ttt!]
\begin{center}
\includegraphics[width=0.95\textwidth]{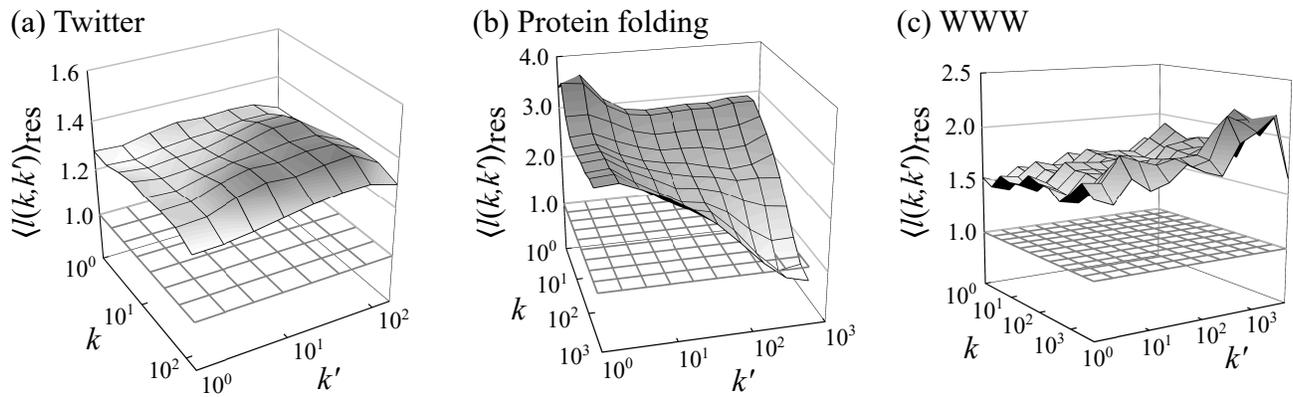}
\caption{Rescaled average shortest path distance
$\langle l(k,k')\rangle_{\text{res}}$ between two nodes of degrees
$k$ and $k'$ for (a) the Twitter network, (b) protein folding network,
and (c) WWW. Two horizontal axes represent $k$ and $k'$ in logarithmic
scales. The lower flat meshed planes indicate the rescaled average path
distances for NNCRNs.}
\label{fig:5}
\end{center}
\end{figure*}
In order to understand more deeply how degrees correlate in a network
with intrinsic LRDC, we measure the average shortest path distance
between two nodes of degrees $k$ and $k'$ defined by Eq.~(\ref{lavkk}).
It is convenient to introduce the rescaled average shortest path
distance $\langle l(k,k')\rangle_{\text{res}}$ defined by
\begin{equation}
\langle l(k,k') \rangle_{\text{res}}=\frac{\sum_{l}l\tilde{P}(l|k,k')}{\sum_{l}l\tilde{P}_{1}(l|k,k')},
\label{reslavkk}
\end{equation}
where the denominator is the average path distance for the
corresponding NNCRN. Figure \ref{fig:5} shows the results for three
real-world networks with intrinsic LRDCs, the Twitter network
\cite{url1}, protein folding network \cite{url4}, and WWW in the domain
of the University of Notre Dame \cite{url1}, which are listed in Table
\ref{table:2}. The results of $\langle l(k,k') \rangle_{\text{res}}$
for these networks are, in common, larger than those for the
corresponding NNCRNs (flat meshed planes). This implies that the
average path distances of the original networks are on average enlarged
by their intrinsic LRDCs. Concerning the Twitter network,
$\langle l(k,k') \rangle_{\text{res}}$ is evenly enlarged almost
independently of $k$ and $k'$. In contrast, for the protein folding
network, the increase of $\langle l(k,k') \rangle_{\text{res}}$ for
low $k$ and $k'$ is much larger than that for high degrees. This
implies that there exists long-range repulsive correlation between low
degree nodes in the protein folding network. On the other hand,
$\langle l(k,k')\rangle_{\text{res}}$ for the WWW is notably enlarged
at high $k$ and $k'$. This shows that the WWW has long-range repulsive
correlation between high degree nodes. As demonstrated above, we can
obtain detailed information on LRDCs in complex networks by comparing
indices defined from the probability distributions with those for
corresponding random networks and NNCRNs.

\section{Conclusions}
\label{sec:conclusion} We have presented a method to distinguish
extrinsic long-range degree correlations (LRDCs) induced by
nearest-neighbor degree correlation (NNDC) and intrinsic ones
attributed to other factors. In a previous work, it has been shown that
the existence of LRDC in a given network $G$ is confirmed by comparing
the probability distributions describing the LRDC in $G$ with those for
corresponding random networks which have the same number of nodes $N$
and the same degree distribution $P(k)$ as $G$ \cite{Fujiki18}.
Similarly, whether LRDC is extrinsic or intrinsic can be identified by
comparing the probability distributions with those for corresponding
nearest-neighbor correlated random networks (NNCRNs) which are
maximally randomized under the constraint of the same $N$ and NNDC as
$G$. We have given a way to calculate the probability distributions for
an NNCRN with a specific $N$ and NNDC within the mean-field
approximation. The analytical results agree quite well with numerically
calculated ones. Finally, we applied our method to several real-world
networks and found that most of networks in the real world are
intrinsically long-range correlated though there exist uncorrelated and
extrinsically long-range correlated networks. Furthermore, we
demonstrated that some indices defined from the probability
distributions are useful to interpret how degrees correlate in a
network with intrinsic LRDC.

It is well known that various properties of complex networks are deeply
related to NNDCs. For instance, the robustness \cite{Goltsev08} and
synchronizability of networks are affected by NNDC \cite{Motter05}, and
fractal scale-free networks show nearest-neighbor disassortative mixing
by degree. We need to elucidate how these properties are related to
intrinsic LRDCs. In particular, high degree (hub) nodes in fractal
scale-free networks are empirically known to repel each other at long
distance. Thus, the relation of such a long-range repulsive correlation
between hubs to the fractal property must be clarified. Furthermore, it
is also quite interesting to study the influence of LRDCs on the
percolation problem or random-walk process on a correlated network,
because fractal networks are fragile against node elimination and show
anomalously slow diffusion. Such studies require efficient indices to
extract specific aspects of intrinsic LRDCs. We have employed, in this
paper, the average degree of the $l$th neighbor nodes $k_{l}(k)$ and
the rescaled average shortest path distance $\langle
l(k,k')\rangle_{\text{res}}$ to characterize intrinsic LRDCs. The
distribution distance $d(P,P')$ is also useful to quantify the strength
of intrinsic LRDC. One can define many other useful indices from the
joint and conditional probability distributions. By utilizing these
indices, we can understand precisely the relation between network
properties and LRDCs.

\begin{acknowledgements}
The authors thank S.~Mizutaka, T.~Hasegawa, and T.~Takaguchi for
fruitful discussions. This work was supported by a Grant-in-Aid for
Scientific Research (No.~16K05466) and a Grant-in-Aid for JSPS Fellows
(No.~18J20874) from the Japan Society for the Promotion of Science.
\end{acknowledgements}

\end{document}